\documentclass[aps,pra,twocolumn,showpacs,floatfix,superscriptaddress]{revtex4-1}

\usepackage[utf8]{inputenc}
\usepackage{amsmath}
\usepackage{amssymb}
\usepackage{graphicx}
\usepackage{epstopdf}
\usepackage{color}
\usepackage[colorlinks, linkcolor=blue, citecolor=blue, urlcolor=blue,breaklinks=true]{hyperref}
\usepackage[usenames,dvipsnames]{xcolor}
\usepackage{changes}
\usepackage{todonotes}

\newcommand{\cre}[1]{c^\dagger_{#1}}
\newcommand{\des}[1]{c^{}_{#1}}

\begin{document}

\title{Competing valence bond and symmetry breaking Mott states of spin-3/2 fermions on a honeycomb lattice}
\author{D. Jakab}
\affiliation{Institute for Solid State Physics and Optics - Wigner Research Centre for Physics, Hungarian Academy of Sciences, H-1525 Budapest P.O. Box 49, Hungary}
\author{E. Szirmai}
\affiliation{BME-MTA Exotic Quantum Phases Research Group, Institute of Physics, Budapest University of Technology and Economics, Budafoki \'ut 8., H-1111 Budapest, Hungary}
\author{M. Lewenstein}
\affiliation{ICFO-Institut de Ci\`encies Fot\`oniques, The Barcelona Institute of Science
and Technology, Av. C.F. Gauss 3, 08860 Castelldefels (Barcelona), Spain}
\affiliation{ICREA-Instituci\'o Catalana de Recerca i Estudis Avan\c cats, Lluis Companys 23, 08010 Barcelona, Spain}
\author{G. Szirmai}
\affiliation{Institute for Solid State Physics and Optics - Wigner Research Centre for Physics, Hungarian Academy of Sciences, H-1525 Budapest P.O. Box 49, Hungary}

\begin{abstract}
We investigate magnetic properties of strongly interacting four component spin-3/2 ultracold fermionic atoms in the Mott insulator limit with one particle per site in an optical lattice with honeycomb symmetry. In this limit, atomic tunneling is virtual, and only the atomic spins can exchange. We find a competition between symmetry breaking and liquid like disordered phases. Particularly interesting are valence bond states with bond centered magnetizations, situated between the ferromagnetic and conventional valence bond phases. In the framework of a mean-field theory, we calculate the phase diagram and identify an experimentally relevant parameter region where a homogeneous SU(4) symmetric Affleck-Kennedy-Lieb-Tasaki-like valence bond state is present.
\end{abstract}

\maketitle

\section{Introduction}
\label{sec:intro} Mott insulators with antiferromagnetic spin
correlations are in the  center of interest in condensed matter
physics due to their relation to high-$T_c$ superconductivity
 \cite{lee2006doping,anderson1973resonating,fazekas1974on}. They are also intensively
 investigated in quantum information, because of  their
potential applicability for quantum computing
\cite{nielsen2000quantum}. In particular, these systems can be
used to realize states required for measurement-based quantum
computation (MBQC)
\cite{raussendorf2001one,verstraete2004valence,van2006universal,gross2007novel,briegel2009measurement}.
In addition they may exhibit nontrivial
topology~\cite{yang93a,bauer2014chiral,nasu15a,hickey2015haldane}; accordingly, they
play an important role in the studies of the topological states of
matter.  For experimental studies of these phenomena ultracold
atomic systems provide one of the most promising and efficient
playgrounds. There is a still ongoing progress in ground breaking
experiments with ultracold atoms in order to realize quantum
emulators of magnetic systems \cite{lewenstein2012ultracold}. The
main advantage is that in these systems there is an unprecedented
control over the parameters describing almost every feature of
their physics
\cite{jordens2010quantitative,trotzky2010controlling,greif2011probing,nascimbene2012experimental,fukuhara2013quantum,fukuhara2013microscopic,
imriska2014thermodynamics,*imriska2014erratum}. The atoms are
trapped optically; both the potential height and the lattice
periodicity can be adjusted by tuning the amplitude, phase and
wavelength of the lasers. Even the geometry of the lattice can be
changed in situ \cite{tarruell2012creating}. Further advantage of
such systems is that interaction between the atoms can be
controlled in a wide range through the access of various
scattering resonances
\cite{inouye1998observation,fedichev1996influence,petrov2001interatomic,chin2010feshbach}.
As a result of this versatility, many antiferromagnets, either
encountered in real materials, or proposed by theorists for
academic interest, can now be potentially realized
\cite{hermele2009mott,gorshkov2010two,szirmai2011exotic,szirmai2011gauge,sinkovicz2013spin,hazzard2012high,pinheiro2013xyz,cai2013pomeranchuk,cai2013quantum,
cazalilla2014ultracold,grass2014quantum}. In the first experiments
the Mott insulator state was realized with a dilute gas sample of
alkaline atoms loaded to an optical lattice formed by
counterpropagating laser beams
\cite{greiner2002quantum,jordens2008mott,greif2014short}. Later,
trapping of higher spin alkalies
\cite{krauser2012coherent,Pagano14} and cooling of
alkaline-earth-metal atoms to quantum degeneracy
\cite{taie2010realization,desalvo2010degenerate} has opened the
way to Mott insulators with higher spin atoms
\cite{taie2012su,scazza2014observation}. Most recently, the first
steps were made towards the study of the direct effects of
spin-exchange interactions~\cite{Zhang14a,Cappellini14a,Scazza14a}

The word 'antiferromagnet' refers to a state of matter, where the
total magnetization of the sample is zero, but magnetic
correlations differ from that of the trivial paramagnetic phase.
The simplest possible antiferromagnetic state has N\'eel order: in
a square lattice in 2D, for instance, opposing spins are arranged
in a checkerboard configuration. Such a state is symmetry
breaking, and an order parameter can be introduced, which is the
magnetization of the sublattice formed by every second site.
According to the Mermin-Wagner-Hohenberg theorem
\cite{mermin1966absence,hohenberg1967existence} symmetry breaking
cannot take place in two dimensions at finite temperature, while
in one dimension fluctuations destroy long-range order even at
$T=0$ \cite{sachdev2011quantum}. Therefore, another completely
different antiferromagnetic state can be expected in low
dimensions: the singlet covering of the lattice, where pairs of
sites form a two-particle spin-singlet state
\cite{fazekas1999lecture}. Similar state was discovered by
Majumdar and Ghosh in a 1D model \cite{majumdar1969on}, where the
ground state is a periodic lattice of independent singlet pairs. A
translational invariant generalization of such a
valence-bond-solid (VBS) state was introduced by Anderson
\cite{anderson1973resonating} to describe frustrated
antiferromagnetism on a triangular lattice; it is called the
resonating valence bond (RVB) state.

The direct consequences of an antiferromagnetic spin exchange can
be studied most clearly in the Mott insulator state of matter
\cite{lee2006doping,imada1998metal,
fisher1999mott,auerbach1994interacting,fazekas1999lecture}. In
this case the charge degree of freedom is frozen out --- e.g. as a
consequence of a strong repulsive interaction --- and the low
energy physics of the system is governed by the spin degrees of
freedom, which in turn are governed by an effective
Heisenberg-like exchange interactions. However, as mentioned
above, it is not obvious, particularly in 2D, that even in the
case of strong antiferromagnetic exchange interactions, a
N\'eel-type spin ordered state will emerge. Instead, valence bonds
can form and the SU(2) spin rotational invariance can remain
preserved. Nowadays the concept of the valence bond picture is the
most commonly used to describe the underlying order of the various
spin liquid states. With the help of the valence bond picture a
series of phenomena with nontrivial magnetic origin can be
described like various exotic VBS states \cite{wu06a,
hermele2009mott, szirmai2011exotic, hermele11a, szirmai2011gauge,ganesh2011quantum},
or even chiral spin liquid (CSL) states with non-trivial topology
\cite{hermele2009mott, szirmai2011gauge, hermele11a,
sinkovicz2013spin,hickey2015haldane}.
In Ref.~\cite{szirmai14a} it was shown that a topological
charge-Haldane state can emerge on a spin-3/2 fermionic ladder.
The SU(4) symmetric spin-3/2 system has been studied intensively
in the last few years \cite{wu06a, szirmai2011exotic, corboz11a,
corboz12a, xu08a, hung11a}, mostly on the square lattice. Although
some recent numerical results  suggest that a weak SU(4) symmetry
breaking order can emerge \cite{corboz11a}, it is mostly accepted
that the ground state is a VBS state consisting disconnected RVB
plaquettes. In Ref.~\cite{corboz12a} a pure SU(4) Heisenberg model
was studied on a honeycomb lattice; the authors  found that a
spin-orbital liquid phase can emerge in the system which collapse
into a tetramerized VBS-like state in the presence of next nearest
neighbor exchange \cite{lajko2013tetramerization}.

The states, which can be suitable as universal resource for MBQC,
could be searched among these spin liquid-like states, which
preserve the usual SU(2) spin rotational invariance. The
classification of these states, just as the quest for physical
systems realizing them and harvesting their potential use as
universal resource states, is a big challenge of quantum
information theory. A fundamental requirement is that they have to
be the unique ground states of a Hamiltonian with gapped spectrum
and short-range interactions in order to assure their robustness
\cite{Wei15}.
One of the most promising states is the AKLT-state introduced by
Affleck, Kennedy, Lieb and Tasaki who proposed several models in
'dimension one and more' where the ground state is a unique VBS
\cite{affleck1987rigorous,affleck1988valence}. In the contemporary
technical "slang" these are states that are exactly described by
the Matrix Product States (MPS) or Pair Entangled Projected States
(PEPs) with the lowest possible (nontrivial) {\it local dimension}
(cf. \cite{lewenstein2012ultracold}).

However, the 'parent' Hamiltonians of these idealized states are
not easy to realize as low energy effective Hamiltonian of a
fermionic or bosonic Mott insulator with $SU(2)$ invariant
interactions. Let us remind the reader that when the internal
states of particles correspond to the degenerate manifold of the
hyperfine atomic level $F$, i.e. $|F,m_F\rangle$ of an ultracold
spinor gas, the resulting Hamiltonians, in the absence of symmetry
breaking fields, must be $SU(2)$ symmetric. The corresponding spin
Hamiltonians are given then by powers of the nearest neighbor
Heisenberg interactions $H=\sum_{<i,j>}\sum_k a_k({\bf S}_i\cdot
{\bf S}_j)^k$, where $a_k$ are numbers that depend on scattering
lengths (for a review see for instance \cite{Zawitkowski} for
$F=1,3/2,2,5/2$, and for general Fermi systems c.f.
\cite{Hofstetter,congjunwureview,hermele2009mott}).

The situation for the Mott insulators with one particle per site
might be summarized as described below; the case of Mott
insulators with two or more particles per site is much more
complex (cf. \cite{Zawitkowski}).

\begin{itemize}

\item The original AKLT state was proposed  for a spin $F=1$ in
1D. $F=1$ can $s$-wave collide in the $F_{tot}=0, 2$ channels, and
thus are characterized in general by the two distinct scattering
lengths. The effective Hamiltonian in the super-exchange limit
reads then  $H=\sum_{\left<i,j\right>}\sum^{k=2}_{k=0} a_k({\bf
S}_i\cdot {\bf S}_j)^k$ or equivalently
$H=\sum_{\left<i,j\right>}\sum_{S=0, 2}[ q_S\mathcal{P}_S({\bf
S}_i + {\bf S}_j)]$, where $q_S$ are numbers, and $P_S$ are
projections on the total spin $S$. If one could control the two
scattering lengths independently and arbitrarily, one could
realize the case when $H=\sum_{\left<i,j\right>}\mathcal{P}_2({\bf
S}_i + {\bf S}_j)]$, which in 1D corresponds exactly to the AKLT
case. Unfortunately, such control of scattering lengths nowadays
is hardly possible -- see ref. \cite{Zawitkowski} for details.

\item  The Hamiltonian $H=\sum_{\left<i,j\right>}\sum^{k=2}_{k=0}({\bf S}_i\cdot {\bf
S}_j)^k$ for $F=1$ particles is known as {\it
biquadratic-bilinear} Hamiltonian and at least in 1D   has been
studied very intensively (cf.
\cite{Demler2002,Yip2003,Imambekov,oriol2007,gabriele2011} and
references therein). It is known that the antiferromagnetic regime
exhibits a robust gapped phase that does not break the $SU(2)$,
the celebrated Haldane phase \cite{Haldane1,Haldane2}.

\item For $F=2$ bosons there are three possible $s$-wave
scattering channels and three scattering length, respectively. The
effective Hamiltonian has the form
$H=\sum_{\left<i,j\right>}\sum_{S=0, 2, 4}[ q_S\mathcal{P}_S({\bf
S}_i + {\bf S}_j)]$, and can in principle be reduced to
$H=\sum_{\left<i,j\right>}\mathcal{P}_4({\bf S}_i + {\bf S}_j)]$
by adjusting the scattering lengths. That would correspond to AKLT
model on the square lattice or  the 3D lattice with coordination
number 4. Again, in practice the necessary control of scattering
length is not possible.

\item For $F=3/2$ fermions there are two possible $s$-wave
scattering channels and, consequently, two scattering length,
respectively. The effective Hamiltonian has the form
$H=\sum_{\left<i,j\right>}\sum_{S=0, 2}[ q_S\mathcal{P}_S({\bf
S}_i + {\bf S}_j)]$, and cannot be by any means  reduced to
$H=\sum_{\left<i,j\right>}\mathcal{P}_3({\bf S}_i + {\bf S}_j)]$,
which on the honeycomb lattice in  2D corresponds to the AKLT
model. Similar situation holds for higher half-integer spin.

\item The effective model with $F=3/2$ was studied extensively, and
already few years ago there has been a lot of progress in
understanding the special properties of $F=3/2$ and $F=5/2$
Fermi gases. In spin-3/2 systems with contact interaction, Wu {\it
et al.} realized that a generic $SO(5)$ symmetry exists
\cite{Wu38}. These authors also found novel competing orders
\cite{Wu39,shu2005exact}, suggesting a {\it quartetting} phase and the $s$
-{\it wave quintet} Cooper pairing phase.

\end{itemize}

Let us recapitulate: the parent Hamiltonian of the
two-dimensional generalization of the AKLT-state is a
spin-Hamiltonian of spin-3/2 operators $\mathbf{S}_i$ on a
two-dimensional honeycomb lattice:
\begin{equation}
H=g_3 \sum_{\left<i,j\right>} \mathcal{P}_3 (\mathbf{S}_i + \mathbf{S}_j) ,
\end{equation}
where $\mathcal{P}_3 (\mathbf{S}_i + \mathbf{S}_j)$ projects to
the total spin-3 subspace: $ (\mathbf{S}_i +
\mathbf{S}_j)^2=3(3+1)$. In the next section we will demonstrate
in detail that the effective spin Hamiltonian derived from a usual
Hubbard-like model does not contain the $\mathcal{P}_3$ projector.
Although, in a more general model this term would appear, its role
remains always secondary, unless we assume "all mighty"  control
over the scattering length -- typically this term  is generated by
weaker interactions: as a perturbation it can change the ground
state even drastically, but it will never become dominant.

In this paper we show that a spin-3/2 ultracold fermionic system
loaded into a two-dimensional honeycomb lattice has a ground state
similar to the two-dimensional generalization of the AKLT-state
for an extended parameter range of the coupling constants
describing the on-site fermion-fermion interaction. Our analysis
is based on a mean-field study of the system with a suitable
ansatz to describe the coexistence of or even the competition
between site- and bond-centered spin orders. We find that the
AKLT-like homogeneous state is the lowest energy solution in an
extended experimentally reachable parameter region. This state
competes on one hand with the usual site ordered homogeneous
ferromagnetic, and N\'eel-type antiferromagnetic phases, at
appropriate coupling constant values, while on the other hand we
find competition also with exotic spin-Peierls-like dimerized
orders. Our results, for the SU(4) symmetric point, are in
agreement with the algebraic color liquid state found in Ref.
\cite{corboz12a}.

The paper is organized as follows. In Sec. \ref{sec:model} we start with the generalized Hubbard model describing the four-component Fermi gas on an optical lattice. Then, we briefly summarize the steps needed to derive a superexchange model for the system in the Mott insulator limit with one particle per site. In Sec. \ref{sec:meanfield} a mean-field approximation is applied to the magnetic superexchange model. By analyzing the mean-field solutions we characterize and discuss the possible ground state phases of the model in Sec. \ref{sec:phdiag}. Finally, we conclude and summarize our results in Sec. \ref{sec:sum}.

\section{Model}
\label{sec:model}

We consider a system of ultracold spin-3/2 atoms on an optical lattice with honeycomb structure. The atom can be any of the alkali- or alkali-earth-metal-atomic species with total hyperfine spin-3/2. At low temperatures and for sufficiently deep optical lattices the atoms occupy the lowest band formed by the lowest states of the individual sites, and the system is described by the generalized Hubbard model.
\begin{equation}
\label{eq:hubham}
H=-t\sum_{\langle i,j\rangle,\alpha}\cre{i,\alpha}\des{j,\alpha}+\frac{1}{2}\sum_{i,\alpha,\beta}V^{\alpha,\beta}_{\gamma,\delta}\cre{i,\alpha}\cre{i,\beta}\des{i,\delta} \des{i,\gamma},
\end{equation}
where the first sum runs over nearest-neighbor pairs and spin components $\alpha=\lbrace -3/2,-1/2,1/2,3/2\rbrace$, while the second sum runs over the single sites of the lattice and spin components $\alpha$ and $\beta$. We use the convention that Greek letters denote the $z$-component of the atomic hyperfine spin. From now on, an implicit summation over repeated Greek indices is also assumed. The operators $\cre{i,\alpha}$ and $\des{i,\alpha}$ respectively create and annihilate a single-particle state at site $i$ with spin projection $\alpha$. The spin-dependent on-site interaction is described by the tensor $V^{\alpha,\beta}_{\gamma,\delta}$. The total spin and its $z$-component of the colliding particles are conserved. The most general form of such spin dependence can be expressed with the help of projection matrices that project to the two particle tensor product space with a given total hyperfine spin of the two colliding particles.
\begin{equation}
\label{eq:spinmat}
V^{\alpha,\beta}_{\gamma,\delta}=g_0\,(P_0)^{\alpha,\beta}_{\gamma,\delta}+g_2\, (P_2)^{\alpha,\beta}_{\gamma,\delta}.
\end{equation} 
$P_0$ and $P_2$ are the projectors projecting to total spin 0 and 2, respectively. These operators are antisymmetric under the exchange of their upper or their lower indices, i.e. $(P_{e})^{\alpha,\beta}_{\gamma,\delta}=-(P_{e})^{\beta,\alpha}_{\gamma,\delta}=-(P_{e})^{\alpha,\beta}_{\delta,\gamma}$, where $e=0,2$. We note, that $P_1$ and $P_3$ is missing from the sum, since they are symmetric in the respective spin indices and due to the Pauli principle their contribution cancels in Eq. \eqref{eq:hubham}. The coupling constants $g_0$ and $g_2$ are expressed in the usual way together with the hopping amplitude $t$ with the help of the Wannier-function overlap integrals \cite{altland2010condensed}. 

When the number of atoms is equal to the number of lattice sites, furthermore, the on-site interaction is much larger than the tunneling amplitude, then multiple occupancy becomes energetically costly. In this limit, all of the low energy states have exactly one particle per site. The dynamics restricted to this low energy sector contains only spin fluctuations. The superexchange Hamiltonian, $\tilde H$, governing such a  Mott insulator dynamics can be obtained with the help of perturbation theory. Here we follow the procedure and notations of Ref. \cite{szirmai2011exotic}.
\begin{equation}
\label{eq:effham}
\tilde H=-\sum_{\langle i,j\rangle} \Bigg[ \frac{2t^2}{g_0}(P_0)^{\alpha,\beta}_{\gamma,\delta}+\frac{2t^2}{g_2}(P_2)^{\alpha,\beta}_{\gamma,\delta}\Bigg]\cre{i,\alpha}\des{i,\gamma}\cre{j,\beta}\des{j,\delta}.
\end{equation}
In the following we express the projectors with the help of the SU(2) spin operators. For a spin-3/2 system, the single spin Hilbert space is 4 dimensional, and the spin matrices are $4\times4$ Hermitian matrices. We use the representation, where the single site spin basis vectors are eigenvectors of $F_z$. The spin matrices are
\begin{align*}
F_z&=\left[\begin{array}{c c c c}
\frac{3}{2}&0&0&0\\
0&\frac{1}{2}&0&0\\
0&0&-\frac{1}{2}&0\\
0&0&0&-\frac{3}{2}
\end{array}\right],\quad
F_x=\left[\begin{array}{c c c c}
0&\frac{\sqrt{3}}{2}&0&0\\
\frac{\sqrt{3}}{2}&0&1&0\\
0&1&0&\frac{\sqrt{3}}{2}\\
0&0&\frac{\sqrt{3}}{2}&0
\end{array}\right],\\
F_y&=\left[\begin{array}{c c c c}
0&-i\frac{\sqrt{3}}{2}&0&0\\
i\frac{\sqrt{3}}{2}&0&-i&0\\
0&i&0&-i\frac{\sqrt{3}}{2}\\
0&0&i\frac{\sqrt{3}}{2}&0
\end{array}\right].
\end{align*}
The two spin tensor product space is 16 dimensional. $P_0$ projects to the $S=0$ (spin singlet) subspace, which is 1 dimensional, while $P_2$ projects to the 5 dimensional $S=2$ (quintet) subspace. Therefore, the total antisymmetric sector is 6 dimensional. In the antisymmetric sector these two orthogonal projectors span the whole space, i.e.
\begin{subequations}
\begin{equation}
(P_0)^{\alpha,\beta}_{\gamma,\delta}+(P_2)^{\alpha,\beta}_{\gamma,\delta}=\frac{1}{2}(\delta_{\alpha,\gamma}\delta_{\beta,\delta}-\delta_{\alpha,\delta}\delta_{\beta,\gamma})\equiv(\mathbf{E}^{\text{(as)}})^{\alpha,\beta}_{\gamma,\delta}.
\end{equation}
Here, the notation with the superscript '(as)' is introduced as a shorthand for antisymmetrization. Furthermore \cite{szirmai2011exotic},
\begin{multline}
-\frac{15}{4}(P_0)^{\alpha,\beta}_{\gamma,\delta}-\frac{3}{4}(P_2)^{\alpha,\beta}_{\gamma,\delta}=
\frac{1}{2}(\mathbf{F}_{\alpha,\gamma}\cdot\mathbf{F}_{\beta,\delta}-\mathbf{F}_{\alpha,\delta}\cdot\mathbf{F}_{\beta,\gamma})\\
\equiv(\mathbf{F}_1\cdot\mathbf{F}_2)^{\text{(as)}},
\end{multline}
\end{subequations}
where $\mathbf{F}=(F_x,F_y,F_z)$ is the 3 component vector of the spin-3/2 matrices. With a straightforward calculation one obtains the Hamiltonian
\begin{multline}
    \label{Hsajt}
    \tilde H =\sum_{\langle i,j\rangle}\bigg[a_n (n_i\,n_j+\chi_{i,j}\,\chi^\dagger_{i,j}-n_i)\\
+a_s\bigg(\mathbf{S}_i\mathbf{S}_j+\mathbf{B}_{i,j}\mathbf{B}_{i,j}^\dagger-\frac{15}{4}n_i\bigg)\bigg],
\end{multline}
where we have introduced the following two-fermion operators:
\begin{subequations}
\label{eqs:twoferops}
\begin{align}
n_i&=\cre{i,\alpha}\des{i,\alpha},\\
\chi_{i,j}&=\cre{i,\alpha}\des{j,\alpha}, \label{eq:chi}\\
\mathbf{S}_i&=\mathbf{F}_{\alpha,\beta}\,\cre{i,\alpha}\des{i,\beta},\\
\mathbf{B}_{i,j}&=\mathbf{F}_{\alpha,\beta}\,\cre{i,\alpha}\des{j,\beta}.\label{eq:B}
\end{align}
\end{subequations}
The quantities $n_i$ and $\mathbf{S}_i$ describe the density and spin on site i, respectively. In the Mott insulator state with one particle per site, $n_i\equiv1$ on the whole lattice. The operators $\chi_{i,j}$ and $\mathbf{B}_{i,j}$ are bond operators describing nearest-neighbor correlations; $\chi_{i,j}$ is the SU(4) symmetric part, while $\mathbf{B}_{i,j}$ is for correlations violating the spin rotation symmetry. The coupling constants in the SU(4) symmetric and symmetry-breaking channels are $a_n=-t^2(5g_0-g_2)/(4g_0g_2)$ and $a_s=-t^2(g_0-g_2)/(3g_0g_2)$, respectively. We assume that the quintet coupling constant ($g_2$) never dominates so strongly over the singlet one ($g_0$) that the SU(4) invariant interaction would become attractive, otherwise formation of four-particle composite particles would be also expected which is beyond the applicability of our treatment. Accordingly, the coupling constant $a_n$ is always negative. Contrary, the coupling $a_s$ can both be positive or negative. Accordingly, the spin anisotropic interaction can be tuned from a predominantly ferromagnetic exchange to an antiferromagnetic one.

\section{Mean-field theory}
\label{sec:meanfield} 

In order to describe the system with the help of a mean-field theory, we assume that both the site and bond operators can (but not necessarily do) have a nonzero classical value. These classical values of the operators will be denoted by a horizontal bar above the operator. Furthermore, we assume that the fluctuations around these classical values are small: $0\approx(\bar\chi_{i,j}-\chi_{i,j})(\bar\chi_{i,j}^*-\chi_{i,j}^\dagger)$.  This way, $\chi_{i,j}\chi_{i,j}^\dagger \approx \bar\chi_{i,j}^*\chi_{i,j} + \bar\chi_{i,j} \chi_{i,j}^\dagger-|\chi_{i,j}|^2$. We assume similar relations for $\mathbf{S}_i$ and $\mathbf{B}_{i,j}$.

By dropping the constant terms from the Hamiltonian \eqref{Hsajt} and using the above approximation together with the definitions in Eqs. \eqref{eqs:twoferops} we arrive to the mean-field Hamiltonian,
\begin{multline}
\label{parmaggio}
H_{\text{mf}} = \sum_{\langle i,j\rangle}\bigg[a_n \Big(\bar\chi_{i,j}^*\,\cre{i,\alpha}\des{j,\alpha} + \bar\chi_{i,j}\, \cre{j,\alpha}\des{i,\alpha}-|\chi_{i,j}|^2\Big)\\
+a_s\Big( \bar{\mathbf{S}}_i\cdot\mathbf{F}_{\alpha\beta}\,\cre{j,\alpha}\des{j,\beta} +\bar{\mathbf{S}}_j\cdot\mathbf{F}_{\alpha\beta}\,\cre{i,\alpha}\des{i,\beta}-\bar{\mathbf{S}}_i\cdot\bar{\mathbf{S}}_j\\
+\bar{\mathbf{B}}_{i,j}^*\cdot\mathbf{F}_{\alpha,\beta}\,\cre{i,\alpha}\des{j,\beta}+
\bar{\mathbf{B}}_{i,j}\cdot\mathbf{F}_{\alpha,\beta}\,\cre{j,\alpha}\des{i,\beta}-|\bar{\mathbf{B}}_{i,j}|^2\Big)\bigg]\\
-\sum_i \varphi_i (\cre{i,\alpha}\des{i,\alpha}-1).
\end{multline}
The last term can be regarded as a Lagrange multiplier enforcing the single particle per site constraint. The Hamiltonian is now quadratic in the fermion operators, and therefore, it can be diagonalized directly. Once the spectrum and eigenvectors of the Hamiltonian \eqref{parmaggio} are known, every physical quantity can be calculated. To this end, quasi-particles are introduced in terms of which the Hamiltonian is diagonal. Their occupation number is given by the Fermi-Dirac distribution function. Most importantly, there is a set of self-consistency equations
\begin{subequations}
\label{gorgonzola}
\begin{align}
1=\bar n_i&=\langle \cre{i,\alpha}\des{i,\alpha} \rangle,\\
\bar \chi_{i,j}&=\langle \cre{i,\alpha}\des{j,\alpha} \rangle,\label{eq:chiexp}\\
\bar{\mathbf{S} }_i&=\mathbf{F}_{\alpha\beta}\langle \cre{i,\alpha}\des{i,\beta} \rangle,\label{eq:Sexp}\\
\bar{\mathbf{B} }_{i,j}&=\mathbf{F}_{\alpha\beta}\langle \cre{i,\alpha}\des{j,\beta} \rangle,\label{eq:Bexp}
\end{align}
\end{subequations}
which has to be solved to find the mean fields.

\begin{figure}[tb]
\centering
\includegraphics[width=0.95\columnwidth]{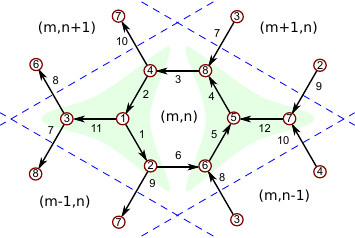}
\caption{(Color online) The choice of the unit cell. There are 8 nonequivalent lattice points inside the unit cell, which are illustrated as numbers inside the circles. There are 8 different bonds running inside the unit cell and 4 bonds connecting sites between neighboring cells. The bonds are directed and are enumerated with numbers. Their orientation is represented by an arrow. The coordinates in parenthesis refer to the position of the unit cell. The shading in the figure is only a guide to the eye, for locating the possible tetramerized clusters.}
\label{fig:unitcell}
\end{figure}
In order to solve the self-consistency equations, one needs to introduce a unit cell repeating periodically on the lattice. The choice of the unit cell is a key assumption in the applied mean-field calculation.
In this system, we expect the competition between symmetry-breaking states with non-vanishing, on-site, classically ordered spins and states with bond-centered mean fields with symmetry breaking or spin-disordered nature. The unit cell has to be compatible with all these assumptions.

States with classically ordered, site-centered spins does not require any special unit cell. In a ferromagnetic state all of the mean-field spins \eqref{eq:Sexp} point to the same direction. In a Néel-like state, the site spins are alternating in a checkerboard manner. Thus on the honeycomb lattice, even the smallest unit cell, containing two sites, can describe both states. The non-zero mean-field value of bond-centered operators signals the presence of spin correlations on clusters of sites. In the spin-3/2 case a completely antisymmetric SU(4) singlet can be formed with 4 sites. In the simplest case, these 4 sites occupy the smallest tetramer, 3 sites surrounding a central one \cite{lajko2013tetramerization}. In order to describe these states too, the unit cell needs to contain all these four sites. If we want to include also competition with states forming larger clusters, we need to introduce a further enlarged unit cell. To this end, we introduce a unit cell containing 8 sites, as depicted in Fig. \ref{fig:unitcell}.

We have $12$ $\bar\chi$ variables and $12\times3$ $\bar{\mathbf{B}}$ ones corresponding to the $12$ bonds, from which 8 is connecting sites inside the unit cell and 4 is linking together neighboring unit cells. These, together with the $8\times3$ spin variables and the $8$ Lagrange multipliers add up to a total of $48$ complex and $32$ real, that is overall $128$ real variables. The solution strategy is the following. The $128$ mean fields and Lagrange multipliers are assumed to be the variables we are looking for. The set of $128$ equations are those in Eqs. \eqref{gorgonzola}. The right hand side of the equations depend on the variables implicitly through the expectation values of the quasiparticles and the eigenvectors of the mean-field Hamiltonian. In the ground state ($T=0$), the quasiparticle occupation is 1 (0) for single-particle states below (above) the Fermi energy. Because of the jump in the occupation numbers at the Fermi energy, the nonlinear solver of the self-consistency equations can get stuck. To find the solutions we first go to finite but low temperatures, where the Fermi-Dirac distribution is more smooth. The occupation numbers are calculated with the help of the thermal average defined by the grand canonical density matrix $\rho=\exp(-\beta H_{\text{mf}})/Z$, with the Hamiltonian \eqref{parmaggio}, $\beta=1/(kT)$ the inverse temperature and $Z=\mathrm{Tr}[\exp(-\beta H_{\text{mf}})]$ the grand canonical partition function. Then, we follow the solution by lowering the temperature to a much lower value. In our calculation the characteristic energy of the problem is $|a_n|$. By choosing a final inverse temperature $\beta=100\times|a_n|^{-1}$ we get results corresponding to the zero temperature limit up to our numeric accuracy.

Even though Eq. \eqref{eq:effham} is derived from the Hubbard Hamiltonian \eqref{eq:hubham}, it has a much higher symmetry. Namely, Eq. \eqref{eq:effham} is invariant under local $\mathrm{U}(1)$ transformations, $\des{j,\alpha}\rightarrow e^{i\theta_j}\,\des{j,\alpha}$, which is not true for Eq. \eqref{eq:hubham}. This emergent gauge symmetry is the consequence of the one particle per site local constraint of the Mott insulating state.
The bond operators are not gauge invariant, they transform as, $\chi_{i,j}\rightarrow  \chi_{i,j}\,e^{-i(\theta_i-\theta_j)}$, and similarly, ${\mathbf{B}}_{i,j} \rightarrow {\mathbf{B}}_{i,j}\,e^{-i(\theta_i-\theta_j)}$.
Consequently, a state of the system is characterized by the full set of the mean fields, $\{\bar \chi_{i,j},\bar{\mathbf{S} }_i,\bar{\mathbf{B} }_{i,j}\}\rightarrow|\Psi_{\text{mf}}^{\bar\chi,\bar{\mathbf{S}},\bar{\mathbf{B}}}\rangle$, but in such a way, that those states are equivalent, whose mean fields are related to each other by a gauge transformation.
Therefore, the states can only be characterized through gauge invariant quantities. The magnitude of the $\bar\chi_{i,j}$ bonds is gauge invariant. Instead of their phase, which is not gauge invariant, we can use the phase $\Phi$ of the Wilson loops $\Pi=|\Pi|e^{i\Phi}$, which are the products of the nonzero $\bar \chi_{i,j}$ bonds. Note, that when a Wilson loop is zero, due to the vanishing of a bond expectation value, one can always choose the bonds to be real along that loop. Inside the periodically repeating unit cell of our choice, there are four distinct elementary plaquettes, whose Wilson loops are:
\begin{subequations}
\label{eqs:WL}
\begin{flalign}
\Pi_1 & =\bar{\chi}_1 \bar{\chi}_2 \bar{\chi}_3 \bar{\chi}_4 \bar{\chi}_5 \bar{\chi}_6 ,\label{eq:WL1}\\
\Pi_2 & =\bar{\chi}_1^* \bar{\chi}_9^* \bar{\chi}_{12}^* \bar{\chi}_4^* \bar{\chi}_7 \bar{\chi}_{11} , \label{eq:WL2}\\
\Pi_3 & =\bar{\chi}_6^* \bar{\chi}_8 \bar{\chi}_7^* \bar{\chi}_3^* \bar{\chi}_{10}^* \bar{\chi}_9 , \label{eq:WL3}\\
\Pi_4 & =\bar{\chi}_5^* \bar{\chi}_{12} \bar{\chi}_{10} \bar{\chi}_2^* \bar{\chi}_{11}^* \bar{\chi}_8^* . \label{eq:WL4}
\end{flalign}
Here the bond indices follow the convention used in Fig.~\ref{fig:unitcell}. 
However, the big Wilson loop encircling all 4 plaquettes is inevitably real. It can be easily checked by taking the product of all 4 Wilson loops Eqs. \eqref{eq:WL1}-\eqref{eq:WL4}. Thus, the phase of the fourth loop is just the opposite of the sum of the previous three.
\end{subequations}
The site spins, $\mathbf{S}_i$, are gauge invariant, and therefore their expectation values are physical quantities. The bond spins, $\mathbf{B}_{i,j}$, on the other hand, are not gauge invariant, and loop operators can not be introduced so straightforwardly. However, we can introduce a bond spin with the following definition:
\begin{equation}
\label{eq:bondspin}
\bar{\mathbf{b}}_{i,j}\equiv\bar{\mathbf{B}}_{i,j}\frac{\bar\chi_{i,j}^*}{|\bar\chi_{i,j}|}.
\end{equation}
The bond spin is gauge invariant by construction, and is the last element needed to characterize the state. In summary, the mean-field solutions are uniquely (up to a gauge transformation) characterized by the absolute values of the 12 $\chi$ bonds, the 8 site spins Eq. \eqref{eq:Sexp}, the 12 bond spins Eqs. \eqref{eq:bondspin} and 3 of the 4 Wilson loops Eqs. \eqref{eqs:WL}.

The equivalence of mean-field states related to each other by gauge transformation is obvious on the physical state vector of the spin system, i.e. on the spin wavefunction \cite{lee2006doping}.
\begin{equation}
\label{eq:Gutzwiller}
\Psi(\alpha_1,\alpha_2,\ldots,\alpha_N)=\Big\langle 0\Big| \des{i_1,\alpha_1} \des{i_2,\alpha_2} \ldots \des{i_N,\alpha_N}\Big|\Psi_{\text{mf}}^{\bar\chi,\bar{\mathbf{S}},\bar{\mathbf{B}}}\Big\rangle.
\end{equation}
The construction of the physical state vector from the mean fields is the so-called Gutzwiller projection. It is easy to see that all of the gauge equivalent states lead to the same physical spin wave function, apart from an unimportant global phase, after Gutzwiller projection \cite{lee2006doping}.

\section{Phase diagram}
\label{sec:phdiag}

\begin{figure}[t]
\centering
\includegraphics[width=0.97\columnwidth]{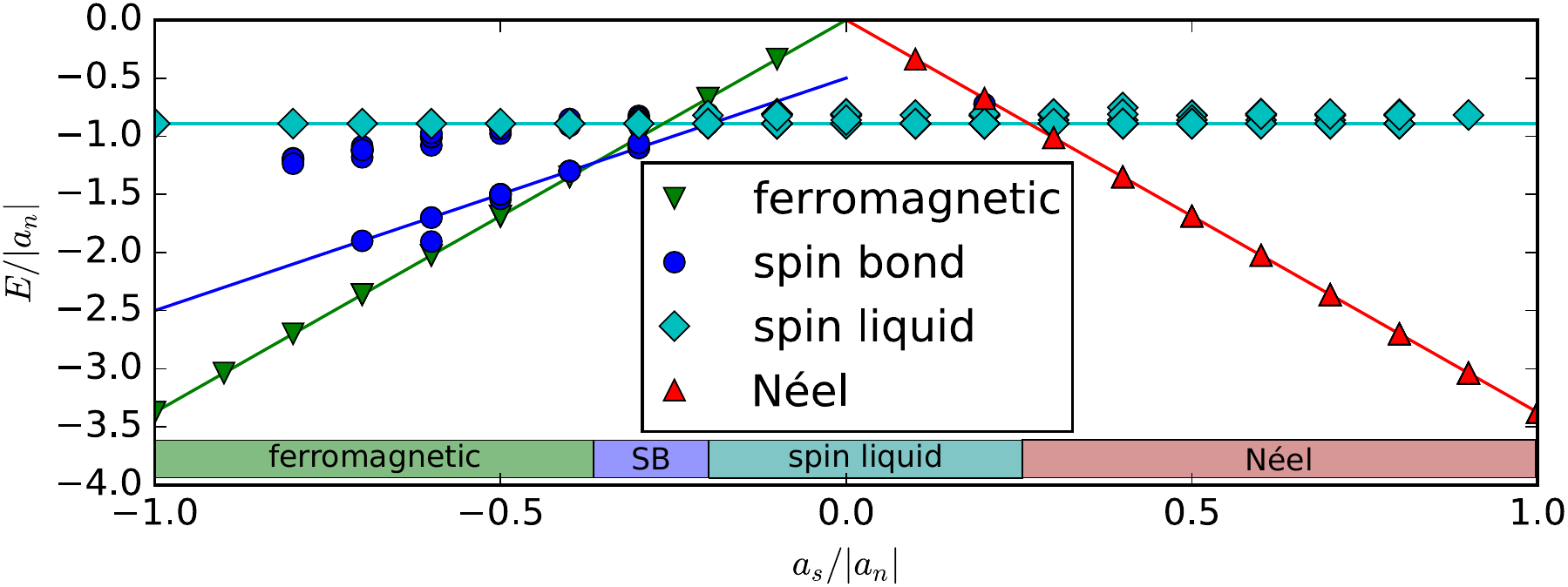}
\caption{(Color online) Energy per site vs coupling strength. On the bottom of the graph, the various phases (SB stands for spin bond) and approximate phase boundaries are indicated. The lines are guides to the eye for the coupling strength dependence of the energy per particle.}
\label{fig:energy}
\end{figure}

In order to obtain the zero temperature phase diagram, we performed a massive search for the solutions of the self-consistency equations \eqref{gorgonzola} numerically, starting from random initial configurations. The phase diagram of the system is determined by the properties of the lowest energy solutions. The energy of various solutions are plotted in Fig. \ref{fig:energy} for different coupling constant ratios. We measure the energy in units of $|a_n|$ (remember that $a_n<0$), and therefore only the ratio $a_s/|a_n|$ is relevant when studying the properties of the ground state.

\subsection{The $\pi$-flux state}

\begin{figure}[t!]
\centering
\includegraphics[]{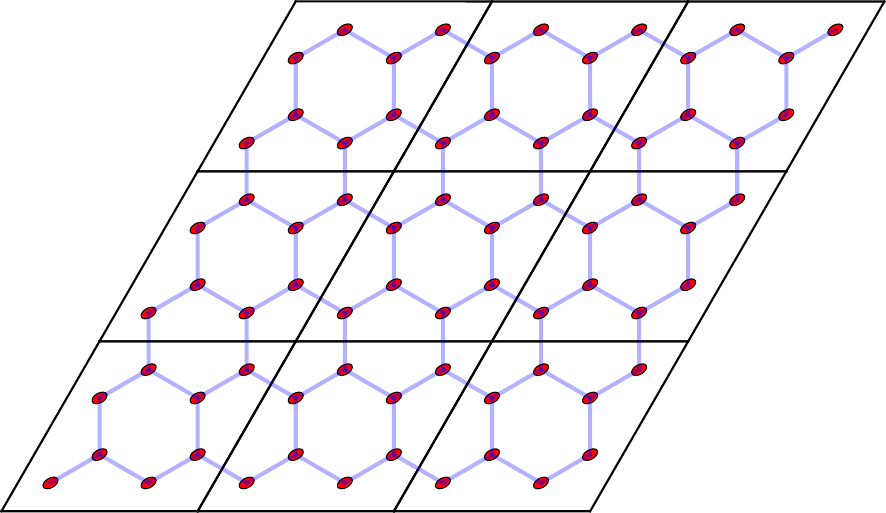}
\caption{(Color online) Illustration of the $\pi$-flux state. The sites are represented by red ellipses, while the homogeneous $\chi_{i,j}$ bonds are drawn with blue lines.}
\label{fig:pifluxstate}
\end{figure}
\begin{figure}[b!]
\centering
\includegraphics[width=.97\columnwidth]{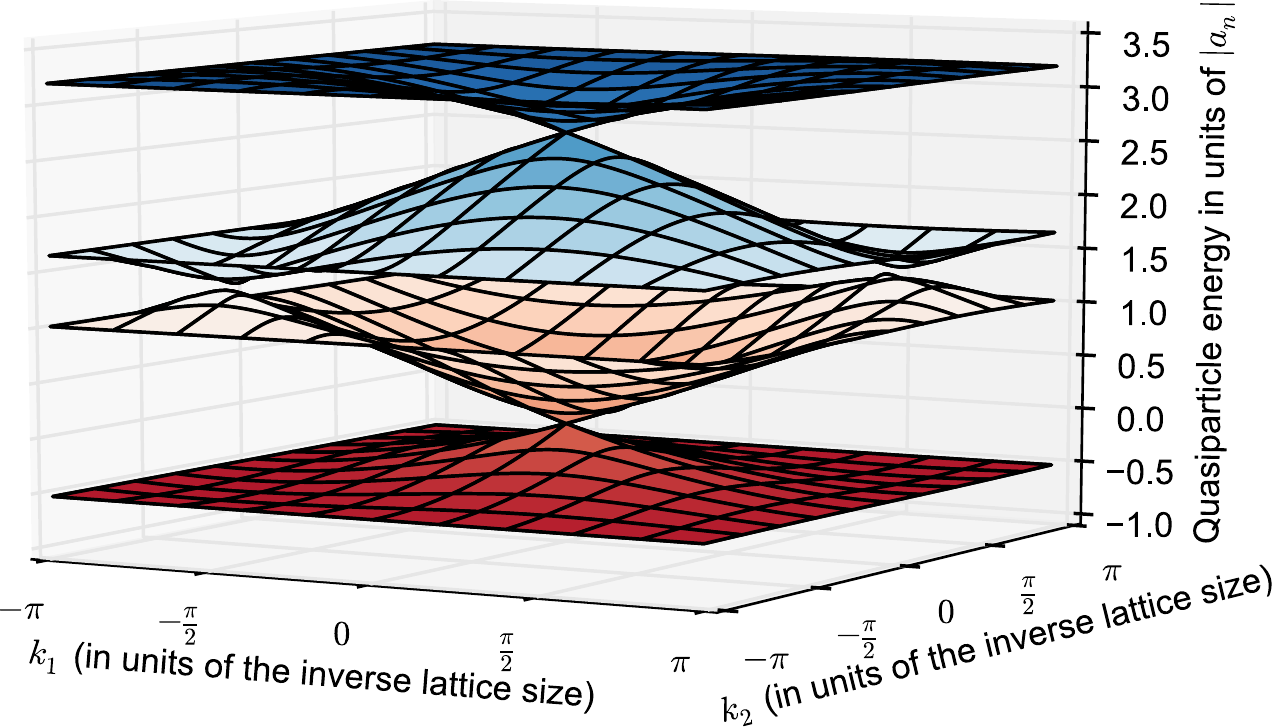}
\caption{(Color online) The fermion energy spectrum of the $\pi$-flux color liquid phase.}
\label{fig:vbsenerg}
\end{figure}
\begin{figure}[t!]
\centering
\includegraphics[width=.97\columnwidth]{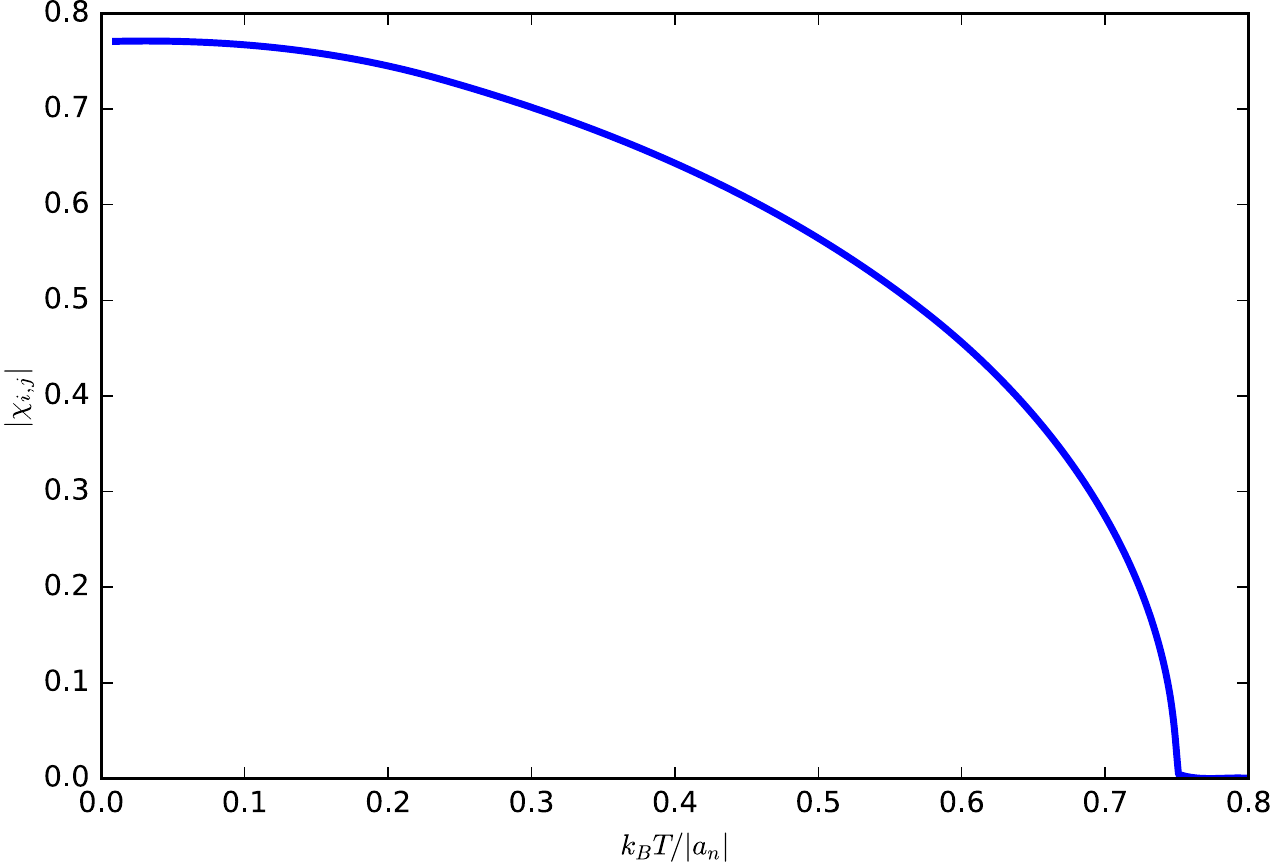}
\caption{(Color online) The temperature dependence of the dimensionless order parameter $|\chi_{i,j}|$ of the $\pi$-flux color-liquid phase at $a_s=0$.}
\label{fig:chivstemp}
\end{figure}

Let us start with the center of the phase diagram, where the couplings in the singlet and quintet channels are equal ($g_0=g_2$), and thus the spin flipping interaction vanishes ($a_s=0$). In this case, the effective Hamiltonian~\eqref{eq:effham} is SU(4) symmetric. In this high symmetry point, all of the low energy mean-field solutions preserve the SU(4) symmetry: there is neither site- nor bond spin order, $\bar{\mathbf{S}}_i=0$, $\bar{\mathbf{B}}_{i,j}=0$. Such a state is exclusively characterized by the plaquette Wilson loops. For the lowest energy solution, we found that all of the SU(4) symmetric bond averages have the same magnitude, $|\bar{\chi}_{i,j}|\approx0.771$, as it is illustrated in Fig.~\ref{fig:pifluxstate}. The Wilson loops are $\Pi_1=\Pi_2=\Pi_3=\Pi_4\approx-0.21$. As the Wilson loops on the elementary plaquettes have a uniform negative value, this state is a homogeneous $\pi$-flux state. The energy of this state is found to be $E/N\approx-0.892|a_n|$ per site. In a $\pi$-flux state, time reversal symmetry is preserved, therefore --- despite the nonzero flux passing through the elementary plaquettes --- this state is nondegenerate, as it is expected from a potential universal resource state. The single fermion excitation spectrum of this homogeneous SU(4) state is shown in Fig.~\ref{fig:vbsenerg}. Due to the complete rotational invariance in the 4 dimensional spin space, and also to the choice of our unit cell, the spectrum consists of 4 bands, each of them are 8-fold degenerate. In our case of 1/4 filling, the lowest band is completely filled. There is a Dirac cone touching between the fully filled lowest band and the empty second band. Therefore, massless Dirac-fermion excitations determine the low energy properties of the homogeneous $\pi$-flux state. We found that this state remains stable at higher temperatures, too, up to a critical temperature around $k_B T_c \approx 0.75|a_n|$. The magnitude of the homogeneous order parameter is plotted versus the temperature in Fig. \ref{fig:chivstemp}. Close to the critical temperature, the order parameter $|\chi_{i,j}|$ disappears at $T_c$ like a square root of the reduced temperature, according to the mean-field exponent $\beta=1/2$, as thermal fluctuations destroy the VBS order.  

We also found that this SU(4) symmetric $\pi$-flux state is robust against the presence of weak SU(4) symmetry violating perturbations, like a nonzero $a_s$. It is particularly remarkable, that despite the Hamiltonian has only SU(2) global symmetry for nonzero $a_s$, the ground state is still invariant under the higher SU(4) symmetry. This symmetry enlargement is unusual, although not unique in high spin systems\cite{assaraf2004dynamical}. Furthermore, we found that in the homogeneous $\pi$-flux state the order parameter $\bar{\chi}_{i,j}$ does not depend on the SU(4) symmetry-breaking coupling $a_s$ at all. As a consequence, the energy of the solution does not depend on $a_s$, and from Eq.~\eqref{parmaggio} its explicit form is given by $E_{\pi\mathrm{flux}}/N= a_n \sum_{\left<i,j\right>} |\bar{\chi}_{i,j}|^2\approx - 0.771^2 |a_n|\cdot 12/8\approx -0.892|a_n|$, where 12 is the number of the independent bonds, 8 is the number of the sites in a unit cell. The energy per particle of the solution is indicated in Fig.~\ref{fig:energy} by a solid line.

The $\pi$-flux state is identical to the one reported as the ground state (algebraic color liquid) of the pure SU(4) Heisenberg model ($a_s=0$) in Ref.~\cite{corboz12a}. Interestingly, in our mean-field calculation the $\chi_{i,j}$ parameters are obtained in a self-consistent way, and even without performing the Gutzwiller projection, the calculated energy of our state ($E/N\approx-0.892|a_n|$) compares remarkably well to the one after the Gutzwiller projection performed with Monte-Carlo calculation: $E/N\approx-0.894|a_n|$ of Ref. \cite{corboz12a}. Note, that the fermion spectrum, Fig. \ref{fig:vbsenerg} also agrees to the one found in Ref. \cite{corboz12a}.

As the strength of the spin flipping interaction $|a_s|$ is increasing, the homogeneous $\pi$-flux phase remains the lowest energy mean-field solution up to the two distinct critical values in the ferromagnetic and Néel sides.  Even beyond the critical value of $a_s$, the homogeneous, SU(4) symmetric $\pi$-flux state remains a higher energy solution above the symmetry-breaking states, although, above $a_s^{c,1}\approx 0.26|a_n|$, the antiferromagnetic order becomes energetically more favorable. On the other side, for negative values of $a_s$, aready above the critical $a_s^{c,2}\approx -0.2|a_n|$ coupling, the ferromagnetic spin exchange starts to favor a symmetry-breaking state. However, before the fully polarized ferromagnetic order wins, in an extended region we found an intermediate state where spin and bond orders coexist, as it will be discussed in Section~\ref{sec:spin-bond}.

\subsection{Conventional symmetry-breaking states}
\begin{figure}[t!]
\centering
\includegraphics[]{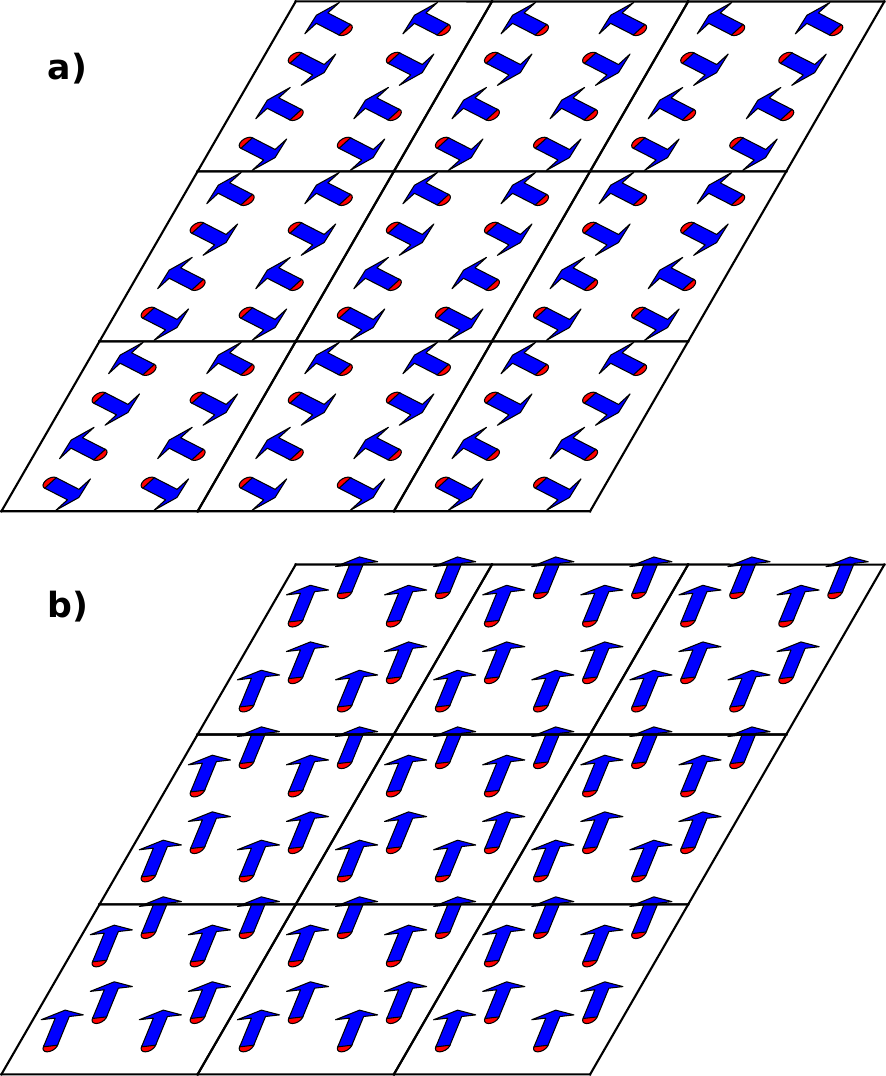}
\caption{(Color online) Illustration of the Néel (a) and ferromagnetic (b) states. The site spins are represented with blue arrows.}
\label{fig:SBstates}
\end{figure}
{\it N\'eel state:} --- In the Néel state both the SU(4) symmetric bond and spin-bond expectation values are zero, $\bar\chi_{i,j}=\bar{\mathbf{B}}_{i,j}=0$. In contrary, the site spins  $\bar{\mathbf{S}}_i$ are nonzero. They have a homogeneous magnitude with maximal spin projection: $\pm$3/2, but spins on neighboring sites are pointing to opposite directions. A nice example of the emerging antiferromagnetic order is shown in Fig.~\ref{fig:SBstates} a). We have seen above, that when the coupling $a_s$ becomes larger than the critical value $a_s^{c,1}\approx 0.26 |a_n|$, the energy of the Néel state goes below that of the $\pi$-flux state. Note, that below $a_s^{c,1}$ iterations of the self-consistency equations starting from random initial mean-field values usually converge to non-symmetry-breaking states. This also indicates the extreme robustness of the SU(4) preserving ground state even in presence of weak $a_s$ couplings. The energy of this classical N\'eel order shows the usual linear dependence on the exchange coupling $a_s$: $E_{\mathrm{Neel}}/N=-a_s (3/2)^3$, as it can be observed in Fig.~\ref{fig:energy}, too.

{\it Ferromagnetic state:} --- For large and negative $a_s/|a_n|$, the spins prefer parallel alignment, due to the strong ferromagnetic coupling dominating the exchange processes. Accordingly, this homogeneous spin ordered state is a usual ferromagnetic state, in which only the spin expectation values $\bar{\mathbf{S} }_i$ are nonzero, and their values are independent of $i$. A typical ferromagnetic state is illustrated in Fig.~\ref{fig:SBstates} b). Note, that we found the ground state to be fully polarized ($|\bar{\mathbf{S} }_i|=3/2$) as soon as the ferromagnetic state becomes the lowest energy state. This happens for $a_s<a_s^{c,3}\approx-0.36|a_n|$. Similarly to the classical N\'eel state, in the fully polarized system the energy dependence on the spin exchange is $E_{\mathrm{FM}}/N= a_s (3/2)^3$. The line corresponding to this energy scaling is also plotted in Fig.~\ref{fig:energy}.

\subsection{The spin-bond ferromagnetic state}
\label{sec:spin-bond}

\begin{figure}[t!]
\centering
\includegraphics[]{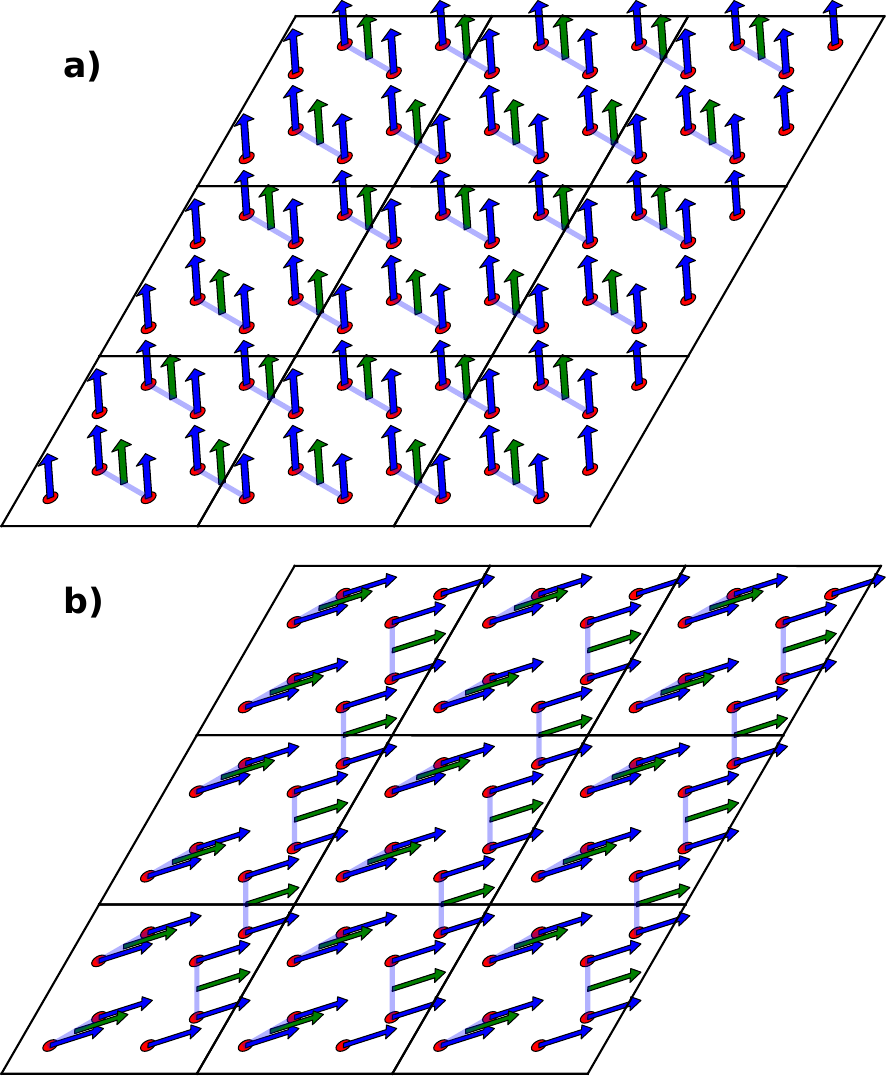}
\caption{(Color online) Illustration of the columnar-bond (a) and zip-bond (b) ferromagnetic state. Note the alignment of the dimers on the two subplots. The site spins are represented with blue arrows, while the bond spins are shown with green arrows.}
\label{fig:spinbondstates}
\end{figure}
On the ferromagnetic side, between the homogeneous $\pi$-flux and fully polarized ferromagnetic phases, the spin anisotropic and the SU(4) symmetric exchange become competitive. In their delicate balance, a mixed intermediate state between $a_s^{c,3}$ and $a_s^{c,2}$ wins, where both spin order and valence bond expectation values on nearest neighbor dimers coexist. Thus, in addition to the homogeneous nonzero spin averages,  both of the two link operators $\chi_{i,j}$ and $\mathbf{B}_{i,j}$ have nonzero expectation values along the dimers. 

The above state, in which a spin anistropic dimer order becomes superimposed upon the ferromagnetic background, has the lowest energy between $a_s^{c,3}$ and $a_s^{c,2}$, i.e. in the vicinity of the $\pi$-flux state. There are two gauge nonequivalent states, which are not related to each other by lattice symmetries either: in one case the dimers form a staggered columnar pattern [Fig.~\ref{fig:spinbondstates} a)], and in the other case, they lay in alternating directions in the neighboring columns forming a zip-like pattern [see Fig.~\ref{fig:spinbondstates} b)]. We found that this state is also robust, the magnitudes of the nonzero order parameters are not sensitive to the tuning of the spin flipping interactions: $|\bar{\mathbf{S} }_i|\approx |\bar{\mathbf{B}}_{i,j}|\approx |\bar{\chi}_{i,j}|\approx 1$. The mean-field energy of the solution from Eq.~\eqref{parmaggio} is $E_{\mathrm{SB}}/N = \sum_{\left<i,j\right>} ( a_n |\bar{\chi}_{i,j}|^2 + a_s |\bar{\mathbf{B}}_{i,j}|^2 + a_s| \bar{\mathbf{S}}_i| |\bar{\mathbf{S}}_j| ) \approx (a_n+4 a_s)/2$, where we used that the number of dimers per unit cell is 4. In Fig.~\ref{fig:energy} the $a_s$ dependence of the energy of the dimer spin-bond solution is also shown with a line. The energy of this dimer spin-bond state goes above that of the ferromagnetic state for $a_s<a_s^{c,3}$, but it still remains a solution, as the magnitude of the ferromagnetic exchange is increasing. Contrary, above the upper critical coupling ($a_s>a_s^{c,2}\approx -0.2 |a_n|$) the SU(4) invariant exchange destroys spin-bond order.

This intermediate spin-bond phase can be regarded as a polarized version of the SU(4)-symmetric dimer phase. To our knowledge, such a state has not yet been found in numerically exact calculations; however, in Ref. \cite{savary2012coulombic}, in the framework of a slave-particle theory, a similar, polarized quantum spin-liquid phase was found.

In the uniform ferromagnetic part of the phase diagram, but not so far from the dimer spin-bond order ($a_s^{c,4}\approx-0.66|a_n|<a_s<a_s^{c,3}$) we found that the competition between the two exchange channels is even stronger leading to various spin-bond states in which the nonzero site spin and bond averages form different complex patterns. The energy of these spin-bond configurations lies lower than that of the dimer spin-bond state. These solutions are not robust, and show a large diversity.

\subsection{Other competing states}
\begin{figure}[t!]
\centering
\includegraphics[]{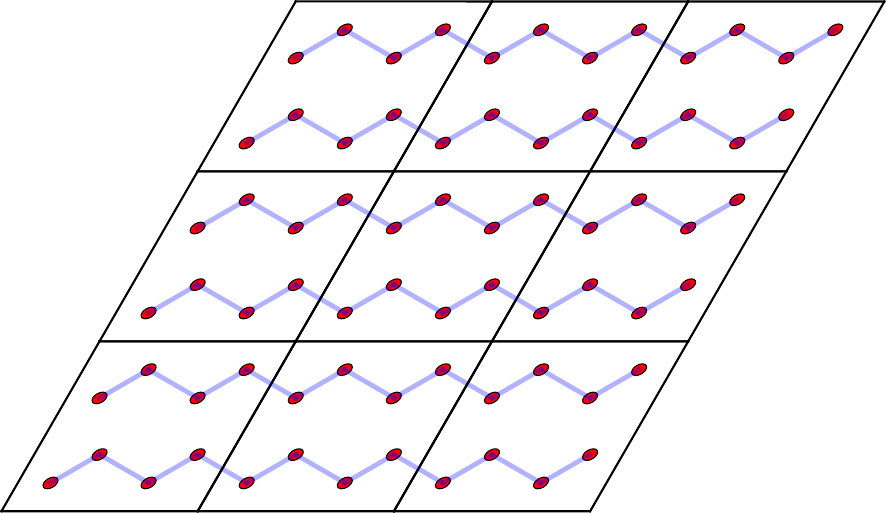}
\caption{(Color online) Illustration of the disconnected-chain state. The $\chi_{i,j}$ bonds are drawn with blue lines. Other mean fields are zero.}
\label{fig:chainstate}
\end{figure}

Above the lowest energy states, we found several other states, some competing with the ground state and some which are situated energetically well above. The competing states located energetically close to the homogeneous $\pi$-flux state in the $a_s^{c,2}<a_s<a_s^{c,1}$ region are especially interesting. These states are all similar to the ground state in many aspects: they are also SU(4) invariant VBS without any spin order. That is,  for these states only the link order parameter $\chi_{i,j}$ has nonzero expectation value, and they are robust against the presence of weak SU(4) symmetry breaking spin flipping processes. However, these states are not homogeneous anymore, usually the translational invariance is preserved in one direction. Therefore these states are three-fold degenerate, corresponding to the lattice symmetry of rotation by $60$ degrees. We mostly found various weakly coupled ladder patterns with stronger bonds along the legs and rungs. The states with different patterns lay energetically close to each other. In Fig.~\ref{fig:chainstate} we present an extreme of such states, where complete dimension reduction can be observed: the bonds along the rungs, and between the ladders are zero, i.e. the nonzero bonds form disconnected chains. The value of the order parameter along the chains is $|\bar{\chi}_{i,j}|\approx 0.897$. As a consequence of the dimension reduction, the fermion spectrum is flat in the reciprocal space along the direction orthogonal to the orientation of the chain.

For $a_s<a_s^{c,3}$, with energies well above the ferromagnetic, and lowest energy spin-bond states, we found several higher energy solutions of the self-consistency equations where the site centered spin order coexists with valence bond order. In these states, the bond spins are disordered and only site spins exhibit the ferromagnetic order.

\section{Summary}
\label{sec:sum}

In this paper we have studied the magnetic properties of Mott insulators realized with ultracold fermions on an optical lattice with honeycomb symmetry. In the framework of a mean-field theory, incorporating both site and bond orderings on equal footing, we calculated the phase diagram and identified the low energy competing magnetic states. We found, that even though the AKLT parent Hamiltonian can not be realized with ultracold atoms, a color liquid state with a $\pi$-flux per plaquette emerges as the ground state for an extended parameter region surrounding the SU(4) symmetric point. This state is characterized by a completely homogeneous set of valence bonds, similar to the AKLT state.

When the spin changing coupling constant is sufficiently large compared to the SU(4) symmetric coupling constant, the color liquid state goes to a symmetry breaking state. In the antiferromagnetic side there is a direct transition to the classical N\'eel order from the homogeneous $\pi$-flux state. Contrary, in the ferromagnetic side of the interaction, the transition to the fully polarized ferromagnetically ordered state goes through an intermediate phase, which is characterized both by site and bond spin orders. In this intermediate phase we found a narrow region where the emerging state can also be regarded as a dimerized VBS state, but with anisotropic nearest-neighbor correlations. For stronger ferromagnetic exchange we found a spin-bond "disordered" state, in which the nonzero links form various disordered patterns as a consequence of the competition between the SU(4) symmetry breaking($a_s$) and preserving ($a_n$) couplings.  When the ferromagnetic coupling is further increased, the system ultimately goes to a ferromagnetic state with only site spin order. 
In the SU(4) symmetric point, our mean-field results are supported by the numerically exact methods of Ref. \cite{corboz12a}, therefore we believe that at the SU(4) point, and also in a vicinity of it, the mean-field result can be trusted. Also, for high values of $|a_s|$, where one anticipates symmetry-breaking solutions, the conclusions of the mean-field theory look solid. Between these different symmetry limiting cases there has to be at least one phase transition. As the mean-field theory is not suitable to describe criticality, it is hard to tell the precise location of the transition point(s). The intermediate spin-bond phase, between the ferromagnetic and color-liquid phases can not be justified, to our knowledge, by earlier numerically exact calculations. However, a similar phase was found in Ref. \cite{savary2012coulombic} for a different model in a slave-particle calculation.

These phases can be distinguished in experiments by measuring quantities sensitive to the type of ordering. The most striking difference between the various phases is exhibited in nearest-neighbor spin correlations, which can be measured e.g. with the superlattice technique \cite{trotzky2010controlling}. Another way of testing the magnetic properties is by measuring the spin structure factor by spin-sensitive Bragg scattering \cite{hart2015observation}. An important question is how low the temperature should be in order to access these correlated phases. The color-liquid phase was found to be robust up to a critical temperature in the order of $|a_n|$, that is a few nanoKelvin in ultracold atom experiments on optical lattices. As for the symmetry breaking phases, we don’t expect them at finite temperatures in an infinite system due to the Mermin-Wagner theorem. However, in experiments in a finite system a tendency towards ordering can happen even in finite temperatures. The zero temperature results can be valid up to the gap of the excitations, which goes approximately with $|a_s|$.

\section*{Acknowledgements}
The authors acknowledge support from the Hungarian Scientific
Research Fund (OTKA) (Grant Nos. PD104652, K105149 and K100908),
Spanish MINECO Project FOQUS (FIS2013-46768), ERC AdG OSYRIS, EU
IP SIQS, EU STREP EQuaM, and EU FETPROACT QUIC. GSz. and ESz. also
acknowledge support from the J\'anos Bolyai Scholarship.

\bibliography{magnetism}

\end{document}